\documentclass[prl, amsfonts, twocolumn, nofootinbib, showpacs]{revtex4}
\usepackage{graphicx,epsfig}

\draft

\newcommand{\be}{\begin{equation}}
\newcommand{\ee}{\end{equation}}
\newcommand{\bea}{\begin{eqnarray}}
\newcommand{\eea}{\end{eqnarray}}

\newcommand{\gapp}{\mathrel{\raise.3ex\hbox{$>$}\mkern-14mu
              \lower0.6ex\hbox{$\sim$}}}
\newcommand{\lapp}{\mathrel{\raise.3ex\hbox{$<$}\mkern-14mu
              \lower0.6ex\hbox{$\sim$}}}
\def\bbox{{\,\lower0.9pt\vbox{\hrule \hbox{\vrule height 0.2 cm
\hskip 0.2 cm \vrule  height 0.2 cm}\hrule}\,}}
\begin{document}

\title{Evaporation of a black hole off of a tense brane}

\author{De-Chang Dai$^1$}
\author{Nemanja Kaloper$^2$}
\author{Glenn D. Starkman$^{1,3}$}
\author{Dejan Stojkovi\'c$^1$}

\affiliation{$^1$Department of Physics, Case Western Reserve
University, Cleveland, OH 44106, USA}

\affiliation{$^2$Department of Physics, University of California,
Davis CA 95616, USA}

\affiliation{$^3$Astrophysics Department, University of Oxford,
Oxford, OX1 3RH, UK}


\begin{abstract}

We calculate the gray-body factors for scalar, vector and graviton
fields in the background of an exact black hole localized on a
tensional 3-brane in a world with two large extra dimensions. Finite
brane tension modifies the standard results for the case with of a
black hole on a brane with negligible tension. For a black hole of a
fixed mass, the power carried away into the bulk diminishes as the
tension increases, because the effective Planck constant, and
therefore entropy of a fixed mass black hole, increase. In this
limit, the semiclassical description of black hole decay becomes
more reliable.
 \end{abstract}

\pacs{04.50.+h,04.60.-m, 11.25.Mj \hfill hep-th/0611184}

\maketitle
\indent

A distinct signature of low scale quantum gravity models
\cite{add} is the possibility of black hole production in high
energy collisions \cite{BHacc}-\cite{AS}. Such processes could be
probed in particle accelerator experiments in the very near
future. Recently a great deal of effort has been expended for the
precise determination of observational signatures of such events.
Chief among them are the black hole production cross section and
the Hawking radiation spectrum. To date, virtually all the work
has been done for the idealized case where the brane tension is
completely negligible. This is because it is generically very hard
to obtain exact solutions of higher-dimensional Einstein's
equations describing black holes on branes with tension. On the other
hand, one generically expects the brane tension to be of the order
of the fundamental energy scale in theory, being determined by the
vacuum energy contributions of brane-localized matter fields.

Recently, a metric describing a black hole located on a
$3$-brane with finite tension, embedded in locally flat
$6$-dimensional ($6D$) spacetime was constructed \cite{KK}. In
spherically symmetric $6D$ Schwarzschild gauge, this metric can be
written as
\begin{eqnarray} \label{KM}
&& ds^{2}=-\left[1-(\frac{r_{h}}{r})^{3}\right]dt^{2}
+\frac{dr^{2}}{1-(\frac{r_{h}}{r})^{3}}+ \\
&& r^{2}\{d\theta^{2}+\sin^{2}\theta \left[d\phi ^{2}+\sin^{2}\phi
(d\chi ^{2}+ B^{2}\sin^{2}\chi d\psi ^{2})\right]\}\nonumber \, .
\end{eqnarray}
The parameter  $B\equiv1-\frac{\lambda}{2\pi M^{4}_{*}}$ measures the
deficit angle about the axis parallel with the $3$-brane, in the
angular direction $\psi$, because the canonically normalized angle
$\psi' = \psi/B$ runs over the interval $[0,2\pi/B]$. Here $\lambda$
is the brane tension, and $M_{*}$ is the fundamental mass scale of
$6D$ gravity, defined as the coefficient of the Ricci scalar in
the $6D$ action $ \int d^6 x \sqrt{g_6} M^4_* R_6/2 \in S_{6D}$.
The black hole horizon is at
\be \label{hor}
r_h = \frac{r_s}{B^{1/3}} \, ,
\ee
where $r_s$ is the conventional $6D$ Schwarzschild radius, given
in terms of the black hole ADM mass
\be \label{schw}
r_s = \left( \frac{1}{4\pi^2} \right)^{\frac{1}{3}}
\frac{1}{M_*}\left( \frac{M_{BH}}{M_*} \right)^{\frac{1}{3}} \, .
\ee
Thus the main effect of the brane tension is to change the
relation between the black hole mass and horizon radius. Then in
terms of the geometric quantities, the metric appears the same as
the $6D$ Schwarzschild solution. However, because of the deficit
angle, at large distances ($r\gg r_h$) the geometry asymptotes to
a conical bulk space.

Following \cite{death,eagid}, we take as the black hole production
cross section its horizon area,
\be \label{cross}
\sigma \sim \pi r_{h}^2  \, .
\ee
It is interesting to explore how this quantity depends on the
brane tension $\lambda$. To do this, one must be careful because
one must first appropriately compactify the asymptotic geometry of
the solution (\ref{KM}), in order to be able to think of it as a
small black hole on a brane in large compact dimensions. In this
case, the volume of the compact dimensions will also depend on the
deficit angle $B$. The precise details depend on the compactification.

One simple case would be to imagine ending the
space far from the black hole (\ref{KM}) on a cylindrically
symmetric $4$-brane \cite{nimalaw} and imposing relfection
symmetry about it, in which case the enclosed $2D$ volume would be
$V^{(2)} \simeq 2\pi L^2/B$, when the black hole is much smaller
than the radial size of the extra dimensions, $r_h \ll L$. Because
the $4D$ Planck mass $M_4$ is determined by the Gauss law equation
\cite{add},
\be \label{gauss}
M_4^2 \sim M_*^4 V^{(2)} \simeq \frac{1}{B} M_*^4 L^2 \, ,
\ee
it is then possible to vary the tension on our $3$-brane, and,
by matching that with a change of the  tension on the cylindrical $4$-brane,
simultaneously hold $M_4$ and $M_*$ fixed. This would  decrease $B$ and
enhance the cross section for a fixed value of $M_*$ and $M_{BH}$,
while the decrease of $V^{(2)}$ from the reduction of the
opening angle of the cone can be compensated by the increase in
the linear size of the extra dimension $L$, at least in principle.

Obviously, there are restrictions to how far one might take these
limits in order to not exceed the sub-millimeter bounds on the
corrections to gravity, which were recently reviewed in
\cite{adelnel}. In general, the situation is even more complicated
because one must carefully consider the details of compactification,
and its topological and geometric aspects, as well as the presence
and deployment of any additional branes in the bulk, to determine
various possible contributions to the Gauss law (\ref{gauss}) to see
how they compare to the tension-dependent contribution of the black
hole-bearing brane. These detailed considerations may also alter the
asymptotic form of the metric (\ref{KM}) far from the black hole, as
they generally are expected to do so in any extra-dimensional
framework. This discussion shows that the precise dependence of the
black hole production cross-section on the brane tension is in
general not simple, but is sensitive to the details of
compactification. For the purposes of our discussion, we can however
ignore this issue, and imagine that we can arrange for a
compactification where $M_*$, $M_4$ and $\lambda$ are freely tunable
parameters. With these assumptions we can treat the solution
(\ref{KM}) as a good approximation for a small black hole on a
brane.

The smoking gun of black hole production in colliders, and hence
of low scale quantum gravity, would be their decay via Hawking
(or rather Hawking-like \cite{Vachaspati:2006ki}) radiation. It
is therefore essential to accurately calculate
emission spectra for fields of various spin emitted by a black
hole, whose precise nature is encoded in the black hole gray-body
factors. Now, from the metric (\ref{KM}) it is clear that the
gray-body factors for the brane-localized fields remain formally
identical to those calculated for the tensionless brane, with the
only difference being in the relationship between the black hole
mass and the horizon size, and therefore the geometric cross
section, as encoded in the Eqs. (\ref{hor}), (\ref{schw}) and
(\ref{cross}).

However, the fields propagating in the bulk will have different
gray-body factors. While in the simplest models with large
dimensions only the fields from the graviton multiplet are taken
to propagate in the bulk, in order to suppress a rapid proton
decay, one needs to physically separate quarks from leptons
\cite{AS}. The simplest way to realize this is to localize quarks
and leptons on different branes in the bulk. Then, since gauge
fields must interact with both quarks and leptons, they must also
be propagate through the section of the bulk between the quark and
lepton branes. When a small black hole resides on one of those
branes, having been created in a collision of matter particles
localized on them, it will emit the gauge fields and its possible
superpartners as bulk fields. Thus it is important to to include
bulk fields in the detailed black hole evolution in order to
develop the tools for the study of more realistic braneworlds.

Let us first consider bulk scalars. For small hot black holes we
can ignore the scalar field masses and simply analyze the equation
of motion of a massless bulk scalar $\Box \Psi  =0$, where $\Psi$
propagates in the background (\ref{KM}).  We can separate this
equation into the radial and angular parts,
\begin{equation}
\frac{1}{r^{4}}\frac{d}{dr}\Bigl(Fr^{4}\frac{d}{dr}R(r) \Bigr)+(\frac{\omega ^{2}}{F}-
\frac{\eta}{r^{2}})R(r)=0 \,  ,
\label{radial}
\end{equation}
and
\be \label{angular}
\Delta_4 Y(\theta,\phi,\chi,\psi) =  - \eta Y(\theta,\phi,\chi,\psi) \,  ,
\ee
where $\Delta_4$  is the Laplacian on the deformed $4$-sphere
$ds^2_4 = d\theta^{2}+\sin^{2}\theta \left[d\phi ^{2}+\sin^{2}\phi
(d\chi ^{2}+ B^{2}\sin^{2}\chi d\psi ^{2})\right]$.

Here $\eta$ is the separation constant,
$F=(1-(\frac{r_{h}}{r})^{3})$, and $B$ measures the deficit angle.
For tensionless brane, $B=1$, the angular equation reduces to the
one for the spherically symmetric scalar field. The solutions can
be written as the expansion in hyper-spherical harmonics with
eigenvalues $\eta =L(L+3)$ \cite{3}. For each fixed $L$ there is
$(2L+3)(L+2)(L+1)/6$ states. When $B \neq 1$, spherical symmetry
is broken, and eigenvalues and degeneracies will change, although
the total number of states will remain the same.

To proceed with the general case $B \ne 1$, we separate the Eq.
(\ref{angular}) further, by writing $Y(\theta,\phi,\chi,\psi)=
\Theta (\theta) \Phi (\phi) \Gamma (\chi) \Xi (\psi)$. This yields
four angular equations
\begin{eqnarray}
\frac{1}{\sin^{3}\theta }\frac{d}{d\theta}
\Bigl(\sin^3{\theta}\frac{d}{d\theta}\Theta \Bigr)
-\frac{\eta _{3}\Theta}{\sin^{2}\theta }=-\eta \Theta \, , \nonumber \\
\frac{1}{\sin^{2} \phi}\frac{d}{d\phi}\Bigl(\sin^{2}\phi
\frac{d}{d\phi}\Phi \Bigr) -\frac{\eta_{2}\Phi}{\sin^{2}\phi}
=-\eta_{3}\Phi \, ,  \nonumber \\
\frac{1}{\sin\chi }\frac{d}{d\chi}\Bigl(\sin\chi \frac{d}{d\chi}\Gamma\Bigr)
-\frac{\eta_{1}\Gamma}{\sin^{2}\chi}=-\eta_{2}\Gamma \, , \nonumber \\
\frac{1}{B^{2} }\frac{d^{2}}{d\psi ^{2}}\Xi=-\eta _{1}\Xi \, .
\label{angular1}
\end{eqnarray}
The solutions will therefore be classified by four quantum
numbers, $(L,l_{1},l_{2},m)$ which determine the four eigenvalues
$(\eta,\eta_3,\eta_2,\eta_1)$ respectively. $L$, $l_{1}$, $l_{2}$
and $m$ are non-negative integers (negative $m$'s will have the
same eigenvalues as positive ones, so we can restrict to $m> 0$
without any loss of generality), and we have the hierarchy $L\geq
l_{1} \geq l_{2} \geq |m|$. In the limit of $B=1$, spherical
symmetry implies  $\eta=L(L+3)$, $\eta_{3}=l_{1}(l_{1}+2)$,
$\eta_{2}=l_{2}(l_{2}+1)$, $\eta_{1}=m^{2}$. In this limit
spherical symmetry implies that the eigenvalue $\eta$ depends only
on the quantum number $L$. However when $B\neq 1$, spherical
symmetry is broken and  $\eta$ depends on two quantum numbers,
i.e. $m$ and $L$. To determine the black hole power, we thus need
to determine degeneracies $N_{L,m}$ in the states labelled by $m$
and $L$. We find
\begin{eqnarray}
m=0 &:& N^{\rm scal}_{L,m}=\frac{(L+2)(L+1)}{2}   \, ,   \\
m\neq  0 &:& N^{\rm scal}_{L,m}=(L-m+2)(L-m+1)    \nonumber \, .
\label{nlm}
\end{eqnarray}

\begin{figure}[ht]
    \centering{
    \includegraphics[width=3.2in]{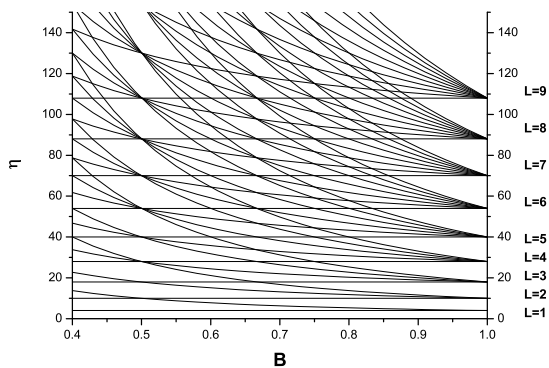} }
    \caption{The figure shows how the eigenvalue $\eta$ changes with $B$.
    $L$ changes from $0$ to $9$.}
    \label{fig:eigen}
\end{figure}

We emphasize here that only the eigenvalue $\eta$ appears in the
radial equation (\ref{radial}). However, because the eigenvalues are
coupled, to find $\eta$ one still needs to solve the full system
of equations (\ref{angular1}).
To determine $\eta $, we adopt the numerical technique
following the Chapter 17.4 of \cite{4}, and rewrite the first three
equations in (\ref{angular1}) as:
\begin{equation}
\frac{1}{\sin^{k}\alpha}\frac{d}{d\alpha}\Bigl(\sin^{k}\alpha
\frac{d}{d\alpha}P\Bigr) +(-\frac{\eta_{k}}{\sin^{2}\alpha}+\eta
_{k+1})P=0 \, . \end{equation}
The parameter $k$ takes the values $k=3,2,1$ and the angle
$\alpha$ stands for $\theta, \phi$ and $\chi$ for the first,
second and third equation in (\ref{angular1}) respectively. After
setting $x= \cos\alpha$ and changing the function $P$ to
$P=(1-x^{2})^{n}T$ with
 $n=(-(k-1)+\sqrt{(k-1)^{2}+4\eta_{k}})/4$, we obtain
\begin{equation} \label{diffeqs}
(1-x^{2})\frac{d^{2}}{dx^{2}}T-(4n+k+1)x\frac{d}{dx}T+(\eta '-2n(2n+k))T=0 \, ,
\end{equation}
with the boundary conditions
\begin{eqnarray} \label{bcs}
x\rightarrow -1&:&\frac{dT}{dx}=-\frac{(\eta '-2n(2n+k))}{(4n+k+1)}T \,  , \nonumber \\
x=0&:&T=0, {\ \ \rm or  \ \ } \frac{dT}{dx}=0 \, . \end{eqnarray}
The two different boundary conditions at $x=0$ correspond to
symmetric (i.e. $dT/dx=0$)  and antisymmetric (i.e. $T=0$)
solutions.
 We can then use the shooting method to solve for the
eigenvalue $\eta$. In Fig. \ref{fig:eigen} one can see how this
eigenvalue changes with $B$ in nine branches of $L$. The $m=0$
branch does not change with $B$. However $m \neq 0$ branches
increase quickly as the brane tension $\lambda$ increases, and so
$B$ decreases. We note here that the increase in the eigenvalue will
result in the reduction of the power of emitted radiation.

\begin{figure}[ht]
    \centering{
    \includegraphics[width=3.2in]{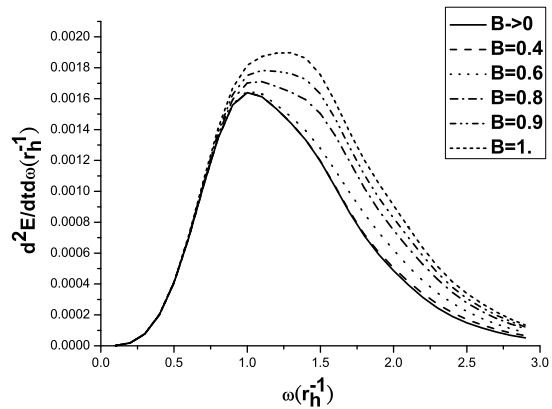} }
    \caption{Hawking spectrum of a $6D$ bulk scalar as a function of brane tension.
 $B$ changes from $B=1$ (tensionless brane) to $B=0$ (deficit angle equal to $2\pi$).}
    \label{fig:scalar}
\end{figure}

Having thus determined the angular eigenvalues $\eta$, we can use
Eq. (\ref{radial}) to find the asymptotic form of the radial
wavefunctions in the  far-field zone and near the horizon, which
are respectively
\begin{eqnarray} \label{asympts}
r\rightarrow r_{h}&:&R^{(h)}=A^{(h)}_{in}e^{-i\omega r^{*}}
+A^{(h)}_{out}e^{i\omega r^{*}} \, , \nonumber \\
r\rightarrow \infty &:&R^{(\infty)}=A^{(\infty)}_{in}
\frac{e^{-i\omega r}}{r^{2}}+A^{(\infty)}_{out}\frac{e^{i\omega r}}{r^{2}} \, .
\end{eqnarray}
Here, $r^*$ is a ``tortoise" coordinate defined by $dr^{*}=dr/F$.
Choosing the boundary condition $A^{(h)}_{out}=0$ which ensures
that near the horizon the solution is purely in-going, we can
numerically integrate Eq. (\ref{radial}) using the forth-order
Runge-Kutta method. From this solution we can calculate the
absorption ratio,
\begin{equation} \label{absorb}
|\tilde{A}_{L,m}|^{2}=1-
\left(\frac{A^{(\infty)}_{out}}{A^{(\infty)}_{in}}\right)^{2} \, ,
\end{equation}
and then, using the principle of detailed balance, write down the
spectrum of Hawking radiation,
\begin{equation} \frac{d^{2}E}{dtd\omega
}=\sum_{L,m}\frac{\omega}{e^{\omega /T_{h}}-1} \frac{N^{\rm
scal}_{L,m}|\tilde{A}_{L,m}|^{2}}{2\pi} \, , \end{equation}
where $T_{h}=\frac{3}{4\pi r_H}$ is the Hawking temperature. In Fig.
\ref{fig:scalar}, we display the results of numerical integration,
which show the variation of the Hawking radiation spectrum with
brane tension, while holding $r_h$ and $M_*$ fixed. For a
tensionless brane, $B=1$, the result coincides with the existing
literature (c.f. in \cite{3}). Increasing the brane tension, and so
decreasing $B$, while assuming that $M_4$ and $M_*$ are held fixed,
reduces the emitted power. We note here that most of the $m\neq 0$
modes have large values for $\eta$ and are thus suppressed with
respect to the $m = 0$ mode, in agreement with the statement that
most of Hawking radiation from a non-rotating black hole is emitted
in the $s$-wave channel \cite{emh} (see also \cite{rotatingBH}).

We should stress that our method of calculating the Hawking
radiance by bulk mode expansion is completely equivalent to first
performing the Kaluza-Klein reduction of bulk scalar perturbations
and then expanding in the $SO(3)$ representations reflecting the
black hole horizon symmetry. This way of thinking about organizing
the calculation is particularly useful in computing the higher
spin emissions, which we turn to next.

The calculation of Hawking radiance for bulk vectors and tensors
goes similarly to the one for scalars, with the exception that one
has to properly account for the helicity multiplets. Namely, in
the case of massless $4D$ vectors and tensors, gauge symmetries
ensure that the only propagating modes are, in the
traceless-transverse Lorentz and de Donder gauge, respectively,
the helicity-1 and helicity-2 modes, like the usual Maxwell and
Einstein fields. However, in the higher-dimensional case they are
accompanied by the Kaluza-Klein towers of massive modes. These
have less gauge symmetry and hence contain more propagating
degrees of freedom, including extra scalars, and scalars and
vectors in the bulk vector and bulk tensor cases, respectively,
which span the full massive $4D$ vector and tensor multiplets
\cite{grw}.

Now, in computing the Hawking radiance from small hot black holes we
can neglect the $4D$ mass for most of these modes since they lie
well below the cutoff while the black hole's temperature is close to
it, but these extra modes still carry away the black hole energy.
Thus we must take them into account. A simple way to do this is to
dimensionally decompose the relevant representations into $4D$
tensors, vectors and scalars, as has been done in \cite{5,6,7}, with
respect to the transverse space to the direction of motion. In
practice, this means counting the tensor representations over the
deformed transverse $S^4$ in Eq. (\ref{KM}), because by the axial
symmetries of the problem the energy will flow radially outward from
the hole. Thus the radial direction and the time can be factored
out. Then, by going to the unitary gauge, we will find that each
degree of freedom obeys a field equation similar to the one for
scalars. Working out the details, one can show that the master
equation for the radial part of a higher dimensional gauge boson or
graviton field (see \cite{5,6,7}) propagating in the background of
the black hole (\ref{KM}) is
\begin{equation}
\frac{d^{2}\Psi}{dr^{*2}}+(\omega^{2}-V(r))\Psi =0 \, ,
\label{master}
\end{equation}
where
\begin{equation}
V(r)=F(\frac{\eta}{r^{2}}+\frac{2}{r^{2}}+\frac{4(1-p^{2})}{r^{5}}) \, .
\end{equation}
The constant p is given by $0$ for scalars and tensor gravitons, $2$
for gravi-vectors, $1/2$ for gauge vectors and $3/2$ for scalar
reductions of bulk vectors. For example, as a quick check we can see
that setting $p=0$ and $\Psi =r^{2}R$ in (\ref{master}) we readily
recover Eq. (\ref{radial}).) One further finds that for the higher
spin fields the angular quantum number $L$ is restricted by the
spin, so that its range is  \cite{5,6,7,8}:
\[L=\left\{ \begin{array}{ll}
             0,1... & \mbox{for scalar perturbation} \,  ,\\
             1,2... & \mbox{for vector perturbation} \, , \\
             2,3... & \mbox{for tensor perturbation} \, .
              \end{array}
             \right.\]
Finally for gravitational scalar perturbations one finds \cite{5,7}
\begin{equation}
V=F\frac{Q(r)}{16r^{2}H(r)^{2}} \, ,
\end{equation}
with
\begin{eqnarray}
Q(r)&=&6400r^{-9}+1920 (\eta -4) r^{-6}-1920 (\eta -4)r^{-3} \nonumber \\
    &+&16(\eta -4)^{3}+96(\eta -4) \,  , \nonumber \\
H(r)&=&(\eta -4)+10r^{-3} \, .
\end{eqnarray}
%


What then remains is to determine the eigenvalues $\eta$ and the
and the number of states for each fixed $\eta$. This number
depends on the helicity, and for the scalars the degeneracy
$N^{\rm scal}_{L,m}$ is the same as in Eq. (\ref{nlm}). For the
vectors, one needs to account properly for gauge fixing when
counting the solutions of the angular equation of motion \cite{8}.
In the end the vector degeneracies depend only on the helicity and
not on how it is lifted into the bulk, yielding
\begin{eqnarray}
m=0 &:& N^{\rm vec}_{L,m}=\frac{L(3L+5)}{2} \\
m\neq  0 &:& N^{\rm vec}_{L,m}=3(L-m+2)(L-m+1) \nonumber \, .
\label{degeneracy2}
\end{eqnarray}
In the case of transverse-traceless tensors, the problem of solving
the gauge fixed angular equation is more difficult. To determine the
degeneracies, we need to solve $\nabla^\alpha \nabla_\alpha T_{\mu
\nu} =0$ satisfying gauge conditions $\nabla^\mu T_{\mu\nu}=0$ and
$T^\mu{}_\mu=0$, where $\nabla^\alpha \nabla_\alpha$ is the
covariant Laplacian on $S^4$, and all the indices $\alpha, \mu, \nu$
run over the deformed transverse sphere of Eq. (\ref{KM}), as we
noted above. Now, below some high (angular!) momentum cutoff, the
total number of states is the same as in the spherically symmetric
limit, computed in \cite{RO} and summarized in the Table I of that
paper. The only effect of broken spherical symmetry is splitting
between some of the levels that were degenerate on $S^4$. Then for
$L \ge |m| \ge 2$ the transverse-traceless tensor can be viewed as
an array of five scalars, and we can simply count up the number of
states for each of them and add them up. This yields $N^{\rm
ten}_{L,m} = 5(L-m+2)(L-m+1) $ for $L \ge | m| \geq 2$. What remains
then is to account for the states with $m=0,1$. We can do this by
taking the total number of states from \cite{RO}, subtracting all
the states with $|m| \ge 2$, whose number we just worked out, and
recalling that the number of states with $L=2, m=0$ is really the
number of independent $s$-wave modes in the full $6D$ graviton
multiplet at some fixed mass level. Since the number of independent
propagating modes of a $6D$ graviton is {\it nine}, we must fix
$N(L=2,m=0) = 9$. This then uniquely determines the degeneracies to
be
\begin{eqnarray}
\label{Nt1}
m=0&:&N^{\rm ten}_{L,m}=\frac{(L-1)(5L+8)}{2} \, , \nonumber \\
m=1&:&N^{\rm ten}_{L,m}=(L-1)(5L+6)\, , \nonumber \\
m\geq 2&:&N^{\rm ten}_{L,m}=5(L-m+2)(L-m+1) \, .
\end{eqnarray}

Then using the same asymptotic form of the radial wavefunctions as
for the scalars (\ref{asympts}) and defining the absorption ratio
for each helicity as $
|\tilde{A^{j}}_{L,m}|^{2}=1-\left({A^{(\infty)}_{out}}/{A^{(\infty)}_{in}}\right)^{2}$,
where $j$ labels the helicity, we can write the Hawking radiance
spectrum for each mode as
\be
\frac{d^{2}E}{dtd\omega }=\sum_{L,m}\frac{\omega}{e^{\omega /T_{h}}-1}
\frac{1}{2\pi}  \sum_j N^{j}_{L,m}|\tilde{A}^{j}_{L,m}|^{2} \, . \label{radiance} \ee
For vectors, $j$ takes values for scalar and Lorentz vector,
whereas for tensors it runs over all scalars, vectors and
transverse-traceless tensor modes with fixed quantum numbers $L$
and $m$. This is the total power emitted in the frequency band
$d\omega$. For the power emitted per degree of freedom, we ought
to divide the result (\ref{radiance}) by four for a gauge boson
and  by nine for graviton, since $6D$ bulk vectors and tensors
have $4$ and $9$ degrees of freedom, respectively, as we have
discussed above.

\begin{figure}[ht]
    \centering{
    \includegraphics[width=3.2in]{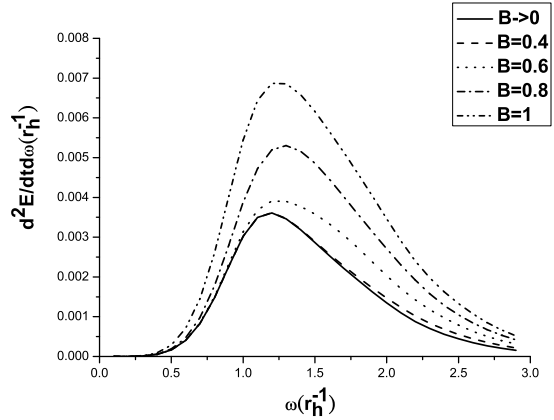} }
    \caption{Hawking radiation spectrum for a bulk gauge boson as a function of the brane tension.
    $B$ changes from $B=1$ (tensionless brane) to $B=0$ (deficit angle equal to $2\pi$).}
    \label{fig:gauge}
\end{figure}

\begin{figure}[ht]
\centering{
\includegraphics[width=3.2in]{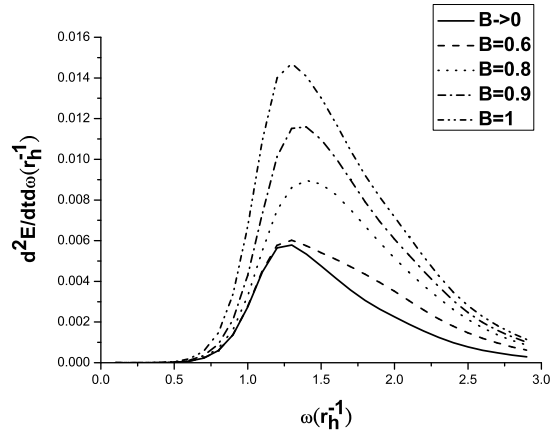} }
\caption{Hawking radiation spectrum for a bulk tensor as a function of the brane tension. Again,
$B$ changes from $B=1$ (tensionless brane) to $B=0$ (deficit angle equal to $2\pi$).}
\label{fig:graviton}
\end{figure}

\begin{figure}[ht]
    \centering{
    \includegraphics[width=3.2in]{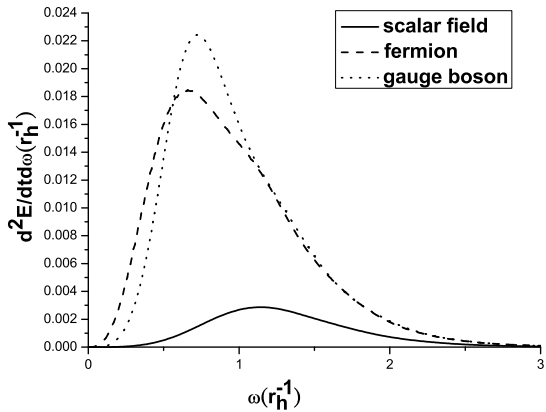} }
    \caption{Hawking radiation of brane localized fields: a scalar,
    a fermion and a vector gauge boson. For fixed horizon size
    they are not affected by the finite
tension of the brane.}
    \label{fig:brane}
\end{figure}

We have computed the power (\ref{radiance}) numerically for the
relevant vector and tensor modes. In Fig. \ref{fig:gauge}  we show
the dependence of the Hawking radiation spectrum for bulk vectors
on the brane tension, while in Fig. \ref{fig:graviton}  we display
the dependence of the radiance on tension for the graviton field.
Finally, for completeness, in Fig. \ref{fig:brane} we show the
Hawking spectrum of brane-localized fields, which for a fixed
horizon size $r_h$ does not depend on the tension. Of course, the
horizon size $r_h$ depends on the tension as given in Eq.
(\ref{hor}). In order to get the power per channel from the total
power displayed in Fig. (\ref{fig:brane}), we need to halve the
fermion and gauge boson distribution.

In sum, collider searches for black holes may be an exciting and
interesting source of information about extra dimensions,
complementing possible astrophysical tests (for recent exploration
of those, see \cite{ruth}). In this note we have explored the
Hawking decay channels for a recently constructed exact black hole
localized on a $3$-brane \cite{KK}. We see that brane tension may
alter significantly the power output of small black holes located on
the brane. While for the non-rotating black holes the dominant
channels are still the brane-localized modes, in realistic models
with split fermions as in \cite{AS}, the gauge boson contribution to
black hole radiance may be significant. Since they need to be in the
bulk to communicate between different (families of) quarks and
leptons, the gauge boson emissions may be a sensitive probe of
deficit angle and therefore brane tension. Specifically, we see that
the power emitted in the bulk diminishes as the tension increases.
This may appear slightly odd at first sight since increasing the
tension for fixed $M_4$ strengthens effective bulk gravity
\cite{KK}. However, when the horizon $r_h$ is held fixed, this comes
about because the gravitational potential barrier depends on the
eigenvalue $\eta$, which as seen in Fig. (\ref{fig:eigen}) increases
as the parameter $B$ decreases, and tension increases. Therefore the
potential barrier grows reducing the rate of energy loss. Similarly,
at fixed black hole mass increasing the tension would simultaneously
increase the horizon size, lowering the temperature and increasing
the black hole entropy. Hence in this case a black hole will also
radiate away its mass more slowly. If we ever produce small black
holes at the LHC, the total Hawking radiance will be a sum of the
power emitting along the brane and off of it. In simple toy models
the emission into the bulk would be subleading because of the
$s$-wave dominance, leading to essentially negligible bulk effects.
However in more realistic models with split fermions and gauge
fields in the bulk, the balance between contributions from the brane
and bulk modes may end up being significantly tilted in favor of the
increasing number of bulk modes. While the precise details of course
would be very model dependent, in such cases the off the wall
contributions may be significant, and so the effects we have
described here may play an important role, warranting a careful and
detailed examination of Hawking radiation losses.

\vskip.2cm {\bf Acknowledgment} \vskip.2cm

NK and GDS thank the Galileo Galilei Institute, Florence, for kind
hospitality in the course of this work. GDS also thanks the
Beecroft Institute for Particle Astrophysics and Cosmology  and
The Queen's College, Oxford for hospitality and support.
The work of NK was
supported in part by the DOE Grant DE-FG03-91ER40674, in part by
the NSF Grant PHY-0332258 and in part by a Research Innovation
Award from the Research Corporation.  The work of GDS was supported
in part by the John Simon Guggenheim Memorial Foundation.

\end{document}